\documentclass[12pt,journal, draftclsnofoot,onecolumn]{IEEEtran}

\usepackage{savesym}
\usepackage{amsfonts,subfigure,multicol,color,verbatim,
graphicx,cite,epsfig,amssymb,amsmath,cases,bm,algorithm,
algorithmic,xcolor,multirow} 
\usepackage{amsmath}
\usepackage{array}
\usepackage{enumerate}
\usepackage{graphicx}
\usepackage{subfigure}

\savesymbol{iint}

\renewcommand{\IEEEQED}{\IEEEQEDopen}
\restoresymbol{TXF}{iint}

\IEEEoverridecommandlockouts

\begin{document}
\markboth{IEEE Wireless Communications} {Peng et al: Cloud
Heterogeneous Radio Access Networks\ldots}

\title{Heterogeneous Cloud Radio Access
Networks: A New Perspective for Enhancing Spectral and Energy
Efficiencies}

\author{
Mugen~Peng,$^{\dagger}$~\IEEEmembership{Senior Member,~IEEE},
Yuan~Li, Jiamo~Jiang, Jian~Li, and
Chonggang~Wang,~\IEEEmembership{Senior Member,~IEEE}

\thanks{Mugen~Peng (e-mail: {\tt pmg@bupt.edu.cn}), Jiamo~Jiang (e-mail: {\tt
jiamo.jiang@gmail.com}), and Jian~Li (e-mail: {\tt
lijian.wspn@gmail.com}) are with the Key Laboratory of Universal
Wireless Communications for Ministry of Education, Beijing
University of Posts and Telecommunications, Beijing, China. Yuan~Li
(e-mail: {\tt liyuansdu@163.com}) is with the Huawei Technologies
Co. Ltd., China. Chonggang~Wang (e-mail: {\tt cgwang@ieee.org}) is
with the InterDigital Communications, King of Prussia, PA, USA.}
\thanks{This work was supported in part by the National Natural Science Foundation of China (Grant No. 61222103), the National High Technology Research and Development Program of China (Grant No. 2014AA01A701), the State Major Science and Technology Special Projects (Grant No. 2013ZX03001001), and the Beijing Natural Science Foundation (Grant No. 4131003).}
\thanks{The paper was submitted on Mar. 22, 2014, and revised on Jul. 17, 2014, accepted on Sep. 08, 2014, and will published in Dec. 2014 in IEEE Wireless Communications on special issue ``Mobile Converged Networks".}}

\date{\today}
\renewcommand{\baselinestretch}{1.5}
\thispagestyle{empty} \maketitle \thispagestyle{empty}
\vspace{-15mm}
\begin{abstract}
To mitigate the severe inter-tier interference and enhance limited
cooperative gains resulting from the constrained and non-ideal
transmissions between adjacent base stations in heterogeneous
networks (HetNets), heterogeneous cloud radio access networks
(H-CRANs) are proposed as cost-efficient potential solutions through
incorporating the cloud computing into HetNets. In this article,
state-of-the-art research achievements and challenges on H-CRANs are
surveyed. In particular, we discuss issues of system architectures,
spectral and energy efficiency performances, and promising key
techniques. A great emphasis is given towards promising key
techniques in H-CRANs to improve both spectral and energy
efficiencies, including cloud computing based coordinated
multi-point transmission and reception, large-scale cooperative
multiple antenna, cloud computing based cooperative radio resource
management, and cloud computing based self-organizing network in the
cloud converging scenarios. The major challenges and open issues in
terms of theoretical performance with stochastic geometry, fronthaul
constrained resource allocation, and standard development that may
block the promotion of H-CRANs are discussed as well.
\end{abstract}


\newpage


\section{Introduction}
Demand for high-speed data applications, such as high-quality
wireless video streaming, social networking and machine-to-machine
communication, has been growing explosively over the past 20 years
and it is envisioned that asymmetric digital subscriber line
(ADSL)-like user experience will be provided in the fifth generation
(5G) wireless systems. This vision implies an average area capacity
of 25Gbps/km$^2$, which is 100 times higher compared with current
fourth generation (4G) systems. Meanwhile, to minimize power
consumptions, a 1000X improvement in the energy efficiency (EE) is
anticipated by 2020. Unfortunately, the cellular network
architecture currently in use is over 40 years old and is not
originally designed for achieving good EE performances, but for the
coverage and mobility consideration. To meet such challenging goals,
revolutionary approaches involving new wireless network
architectures as well as advanced signal processing and networking
technologies are anticipated.

Heterogeneous networks (HetNets) have attracted intense interests
from both academia and
industry\textcolor[rgb]{1.00,0.00,0.00}{\cite{bib:HetNet}}. The low
power nodes (LPN, e.g., pico base station, femto base station, small
cell base station, etc.) is identified as one of key components to
increase capacity of cellular networks in dense areas with high
traffic demands. When traffic is clustered in hotspots, such LPNs
can be combined with high power node (HPN, e.g., macro or micro base
station) to form a HetNet. HetNets have advantages of serving
hotspot customers with high bit rates through deploying dense LPNs,
providing ubiquitous coverage and delivering the overall control
signallings to all user equipments (UEs) through the powerful HPNs.
Actuarially, too dense LPNs will incur the severe interferences,
which restricts performance gains and commercial developments of
HetNets. Therefore, it is critical to control interferences through
advanced signal processing techniques to fully unleash the potential
gains of HetNets. The coordinated multi-point (CoMP) transmission
and reception is presented as one of the most promising techniques
in 4G systems. Unfortunately, CoMP has some disadvantages in real
networks because its performance gain depends heavily on the
backhaul constraints and even degrades with increasing density of
LPNs\textcolor[rgb]{1.00,0.00,0.00}{\cite{bib:CoMP}}. Further, it
was reported that the average spectral efficiency (SE) performance
gains from the uplink CoMP in downtown Dresden field trials was only
about 20 percent with non-ideal backhaul and distributed cooperation
processing located on the base station
in\textcolor[rgb]{1.00,0.00,0.00}{\cite{bib:CoMPgain}}.

To overcome the SE performance degradations and decrease the energy
consumption in dense HetNets, a new paradigm for improving both SE
and energy efficiency (EE) through suppressing inter-tier
interference and enhancing the cooperative processing capabilities
is needed in the practical evolution of HetNets. Meanwhile, cloud
computing technology has emerged as a promising solution for
providing high energy efficiency together with gigabit data rates
across software defined wireless communication networks, in which
the virtualization of communication hardware and software elements
place stress on communication networks and protocols. Consequently,
heterogeneous cloud radio access networks (H-CRANs) are proposed in
this paper as cost-effective potential solutions to alleviating
inter-tier interference and improving cooperative processing gains
in HetNets through combination with cloud computing. The motivation
of H-CRANs is to enhance the capabilities of HPNs with massive
multiple antenna techniques and simplify LPNs through connecting to
a ``signal processing cloud'' with high-speed optical fibers. As
such, the baseband datapath processing as well as the radio resource
control for LPNs are moved to the cloud server so as to take
advantage of cloud computing capabilities. In the proposed H-CRANs,
the cloud computing based cooperation processing and networking
gains are fully exploited, the operating expenses are lowered, and
energy consumptions of the wireless infrastructure are decreased.

\begin{figure}
\centering  \vspace*{0pt}
\includegraphics[scale=1.0]{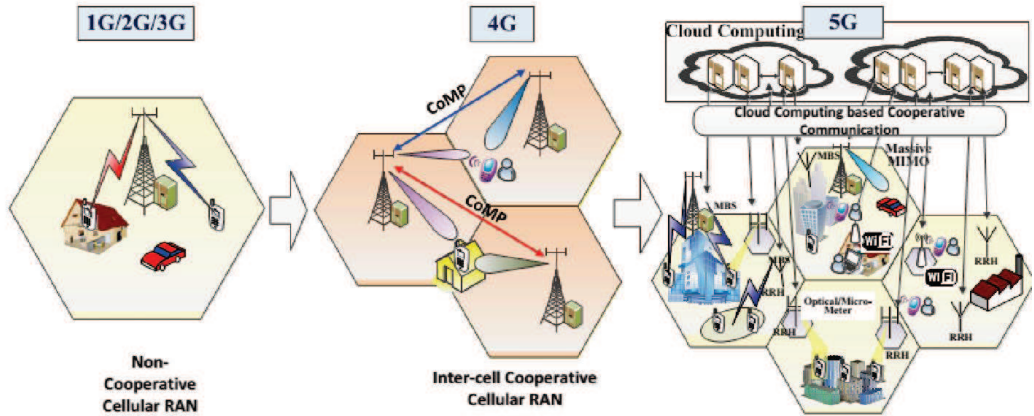}
\setlength{\belowcaptionskip}{-100pt} \caption{\textbf{Cellular
System Evolution to 5G}} \label{fig1}\vspace*{-10pt}
\end{figure}

Fig. \ref{fig1} illustrates the evolution milestone from
conventional 1G to the H-CRAN based 5G system. In the traditional
first, second and third generation (1G, 2G, 3G) cellular systems,
cooperative processing is not demanded because the inter-cell
interference can be avoided by utilizing static frequency planning
or code division multiple access (CDMA) techniques. However, for the
orthogonal frequency division multiplexing (OFDM) based 4G systems,
the inter-cell interference is severe because of the spectrum reuse
in adjacent cells, especially when a HetNet is deployed. Therefore,
the inter-cell or inter-tier cooperative processing through CoMP is
critical in 4G. To control the more and more severe interference and
improve SE and EE performances, the cooperative communication
techniques are evolved from the two-dimensional CoMP to the
three-dimensional large-scale cooperative processing and networking
with cloud computing. Therefore, for the H-CRAN based 5G system, the
cloud computing based cooperative processing and networking
techniques are proposed to tackle aforementioned challenges of 4G
systems and in turn meet performance demands of 5G systems.

In this article, we are motivated to make an effort to offer a
comprehensive survey on technological features and core principles
of H-CRANs. In particular, the system architecture of H-CRANs is
presented, and SE/EE performances of H-CRANs are introduced.
Meanwhile, the cloud computing based cooperative processing and
networking techniques to improve SE and EE performances in H-CRANs
are summarized, including advanced signal processing in the physical
(PHY) layer, cooperative radio resource management (RRM) and
self-organization in the upper layers. The challenging issues
related to techniques and standards are discussed as well.

The remainder of this paper is outlined as follows. H-CRAN
architecture and performance analysis are introduced in Section II.
The promising key technologies related to the cloud computing based
signal processing and networking in H-CRANs are discussed in Section
III. Future challenges are highlighted in Section IV, followed by
the conclusions in Section V.

\section{System Architecture and Performance Analysis of H-CRANs}

\vspace*{15pt}

Although HetNets are good alternatives to provide seamless coverage
and high capacity in 4G systems, there are still two remarkable
challenges to block their commercial developments: \romannumeral1).
The SE performance should be enhanced because the intra- and
inter-cell CoMP need a huge number of signallings in backhaul links
to mitigate interferences amongst HPNs and LPNs, which often results
in the constrained backhaul links; \romannumeral2). The ultra dense
LPNs can improve capacity with the cost of consuming too much
energy, which results in a low EE performance.

Cloud radio access networks (C-RANs) are by now recognized to
curtail the capital and operating expenditures, as well as to
provide a high transmission bit rate with fantastic EE
performances\textcolor[rgb]{1.00,0.00,0.00}{\cite{bib:CRAN}}. The
remote radio heads (RRHs) operate as soft relay by compressing and
forwarding the received signals from UEs to the centralized base
band unit (BBU) pool through the wire/wireless fronthaul links. To
distinguish the advantages of C-RANs, the joint decompression and
decoding schemes are executed in the BBU pool. Accurately, HPNs
should be still critical in C-RANs to guarantee the backward
compatibility with the existing cellular systems and support the
seamless coverage since RRHs are mainly deployed to provide high
capacity in special zones. With the help of HPNs, the multiple
heterogeneous radio networks can be converged, and all system
control signallings are delivered wherein. Consequently, we
incorporate HPNs into C-RANs, and thus H-CRANs are proposed to take
full advantages of both HetNets and C-RAN, in which cloud computing
capabilities are exploited to solve the aforementioned challenges in
HetNets.

\subsection{System Architecture of H-CRANs}

\begin{figure}
\centering  \vspace*{0pt}
\includegraphics[scale=1]{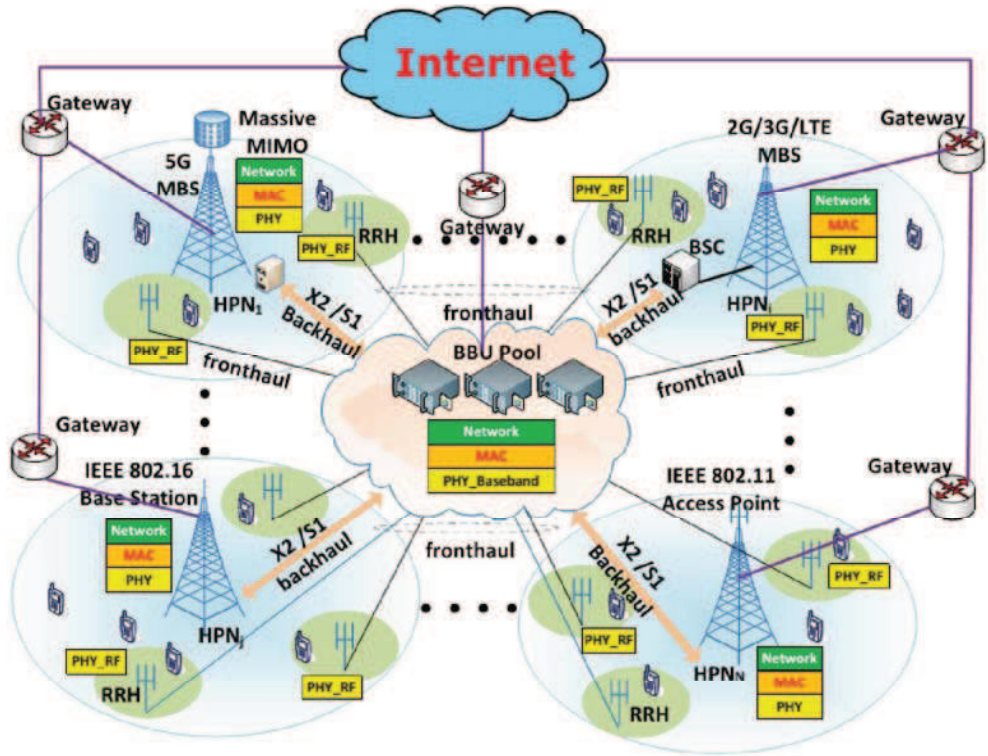}
\setlength{\belowcaptionskip}{-100pt} \caption{\textbf{System
Architecture for Proposed H-CRANs}} \label{fig2}\vspace*{-10pt}
\end{figure}

Similarly with the traditional C-RAN, as shown in Fig. \ref{fig2}, a
huge number of RRHs with low energy consumptions in the proposed
H-CRANs are cooperated with each other in the centralized BBU pool
to achieve high cooperative gains. Only the front radio frequency
(RF) and simple symbol processing functionalities are implemented in
RRHs, while the other important baseband physical processing and
procedures of the upper layers are executed jointly in the BBU pool.
Sequently, only partial functionalities in the PHY layer are
incorporated in RRHs, and the model with these partial
functionalities is denoted as PHY\_RF in Fig. \ref{fig2}.

However, different from C-RAN, the BBU pool in H-CRANs is interfaced
to HPNs for mitigating the cross-tier interferences between RRHs and
HPNs through the centralized cloud computing based cooperative
processing techniques. Further, the data and control interfaces
between the BBU pool and HPNs are added and denoted by S1 and X2,
respectively, whose definitions are inherited from the
standardization definitions of $3^{rd}$ generation partnership
project (3GPP). Since the voice service can be provided efficiently
through the packet switch mode in 4G systems, the proposed H-CRAN
can support both voice and data services simultaneously, and the
voice service is preferred to be administrated by HPNs, while the
high-data packet traffic is mainly served by RRHs.

Compared with the traditional C-RAN architecture, the proposed
H-CRAN alleviates the fronthaul requirements with the participation
of HPNs. Owing to the incorporation of HPNs, the control signallings
and data symbols are decoupled in H-CRANs. All control signallings
and system broadcasting information are delivered by HPNs to UEs,
which simplifies the capacity and time delay constraints in the
fronthaul links between RRHs and the BBU pool, and make RRHs active
or sleep efficiently to save the energy consumption. Further, some
burst traffic or instant messaging service with a small mount of
data can be efficiently supported by HPNs. The adaptive
signaling/control mechanism between connection-oriented and
connectionless is supported in H-CRANs, which can achieve
significant overhead savings in the radio connection/release by
moving away from a pure connection-oriented mechanism. For RRHs,
different transmission technologies in the PHY layer can be utilized
to improve transmission bit rates, such as IEEE 802.11 ac/ad,
millimeter wave, and even optical light. For HPNs, the massive
multiple-input-multiple-output (MIMO) is one potential approach to
extend the coverage and enrich the capacity.

Since all signals are centralized processed in the BBU pool for UEs
associating with RRHs, the cloud computing based cooperative
processing techniques inherited from the virtual MIMO can achieve
high diversity and multiplexing gains. Similarly with C-RANs, the
inter-RRHs interferences can be suppressed by the advanced cloud
computing based large-scale cooperative processing techniques in the
BBU pool. The cross-tier interference among HPNs and RRHs can be
mitigated through the cloud computing based cooperative RRM
(CC-CRRM) via the interface X2 between the BBU pool and HPNs.

To improve EE performances of H-CRANs, the activated RRHs are
adaptive to the traffic volume. When the traffic load is low, some
potential RRHs fall into the sleep mode under adminstration of the
BBU pool. However, when the traffic load becomes tremendous in a
small special zone, both the HPN with  massive MIMO and dense RRHs
work together to meet the huge capacity demands, and even the
corresponding desired RRHs can borrow radio resources from
neighboring RRHs.

\subsection{Spectral and Energy Efficiencies Performances}

By shortening the communication distance between the serving RRH and
desired UEs, and achieving the cooperative processing gains from the
cloud computing in the BBU pool, the SE performance gains are
significant in H-CRANs. Compared with the traditional wireless
cellular networks, multiple RRHs connected to one BBU pool in
H-CRANs could offer much higher performance, where higher degrees of
freedom in the interference control and resource allocation are
achieved. Thanks to the cloud computing technology, the exponential
EE performance gains can be achieved at the cost of linear
increasing SE only when the circuit power is not large in
C-RANs\textcolor[rgb]{1.00,0.00,0.00} {\cite{bib:CRAN}}. Therefore,
the key factor to improve both SE and EE performances is to decrease
the circuit power consumptions of fronthaul links. The efficient
centralized cooling system in the BBU pool and the low transmit
power in RRHs could lead to a significant reduction of the total
energy consumptions. RRHs can be completely switched off to save
much energy when there is no traffic wherein, which presents energy
saving opportunities of approximately 60 percent with contrast to
the non-sleep mode\textcolor[rgb]{1.00,0.00,0.00}{\cite{bib:Sleep}}.
HPNs are responsible for providing the basic service coverage and
delivering the control signallings, while RRHs are used to support
packet traffic with high bit rates. Partial services and overheads
are undertaken by HPNs, which alleviates constraints on fronthaul
and decreases circuit energy consumptions in RRHs, thus improves
both SE and EE performances.

\begin{figure}
\centering
\subfigure[EE Performance Comparisons with various network]{ \label{PC:a} 
\includegraphics[width=2.5in]{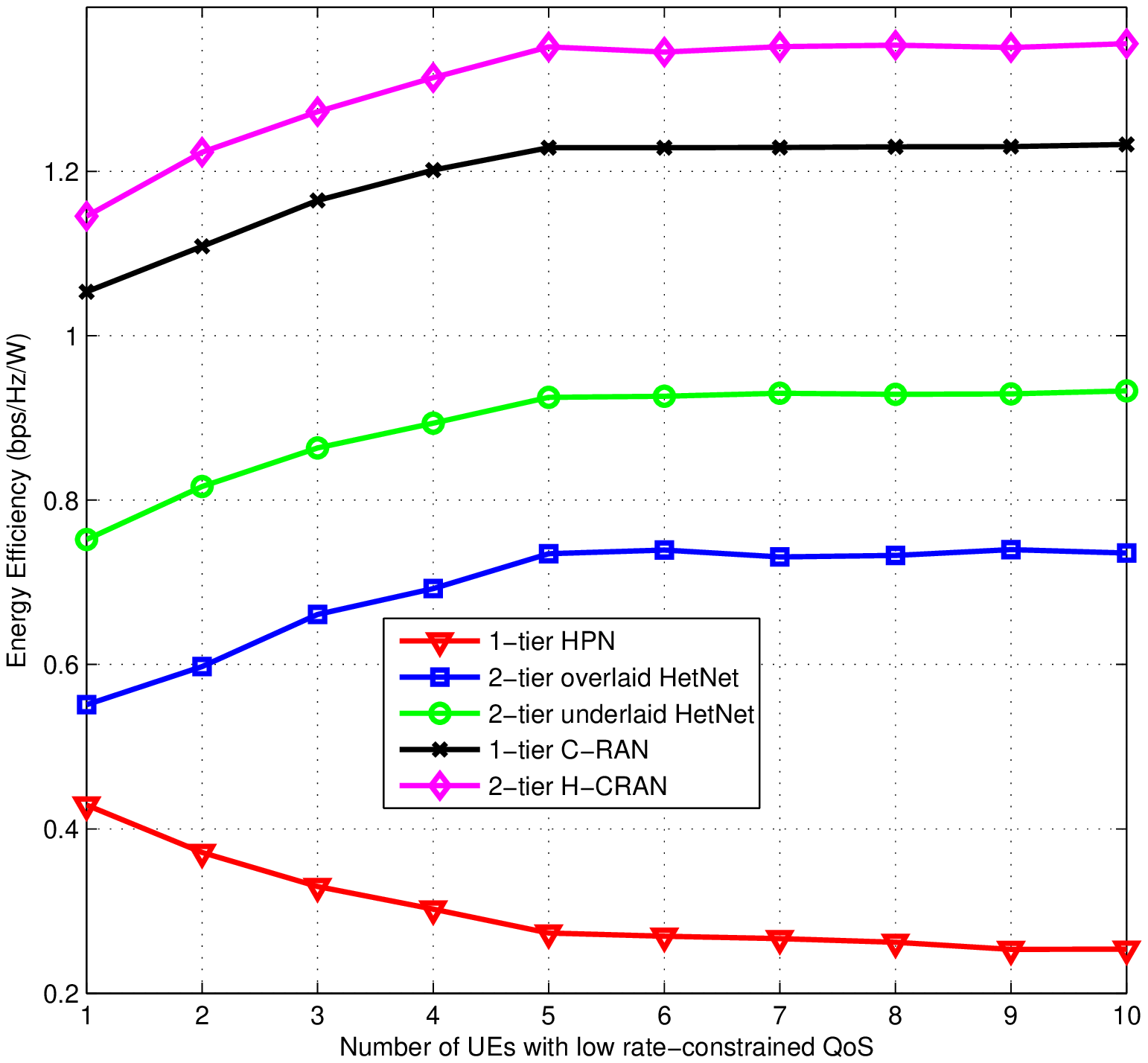}}
\hspace{0.01in}
\subfigure[SE Performance Comparisons with different association number]{ \label{PC:b} 
\includegraphics[width=4.15in]{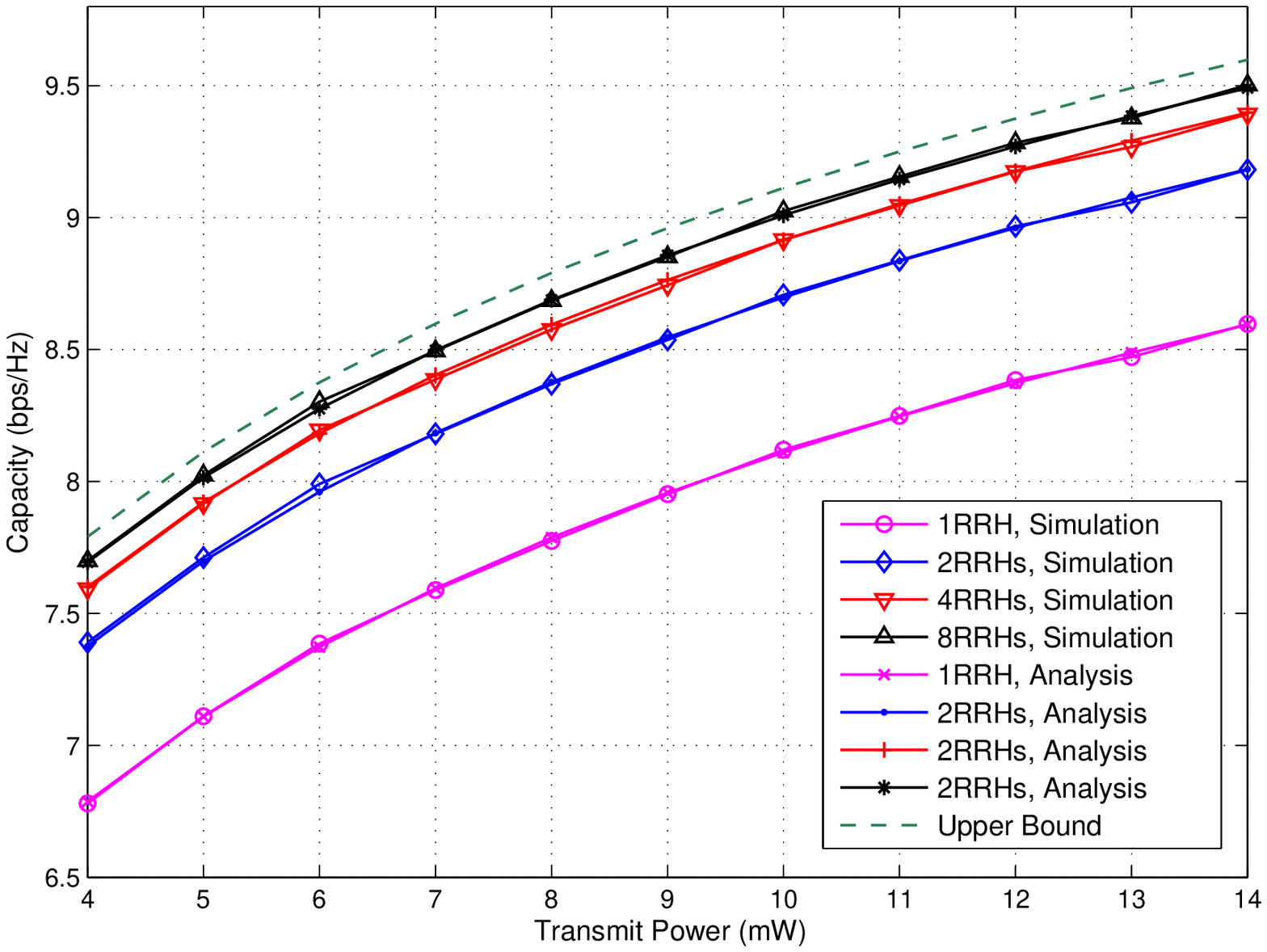}}
\caption{\textbf{EE and SE Performances Evaluations of H-CRANs}}
\label{PC} 
\end{figure}

As shown in Fig. \ref{PC:a}, EE performances in terms of the number
of cell-edge UEs are compared amongst 1-tier HPN, 2-tire underlaid
HetNet, 2-tier overlaid HetNet, 1-tier C-RAN, and 2-tier H-CRAN,
where the same serving coverage, frequency spectrum, transmit power
for both RRH and HPN, number of served UEs are
assumed\textcolor[rgb]{1.00,0.00,0.00}{\cite{peng_TVT}}. The EE
performances decrease with the increasing number of cell-edge UEs in
the 1-tier HPN scenario because more resource blocks and power are
allocated to the cell-edge UEs for guaranteeing their basic
transmission rates. The EE performances are better in 2-tier HetNets
than those in 1-tier HPN because a lower transmit power is needed
and a higher transmission bit rate is achieved. Further, the EE
performances in the 2-tier H-CRAN scenario are better than those in
the 1-tier C-RAN because RRHs in C-RANs suffers from the limited
coverage that HPN could serve. Meanwhile, the EE performances of
1-tier C-RAN are better than those of both 1-tier HPN and 2-tire
HetNet due to gains from the characteristic of cloud computing.

In H-CRANs, there are often ultra dense RRHs in the hot spots, which
results in that a desired UE is associated with multiple RRHs and
HPNs. Therefore, the user-centric RRH/HPN clustering mechanism is
critical, in which the cluster is dynamically optimized for each
active UE, and different clusters for different UEs may overlap. It
is not always optimal that UE associates with the RRH via the
maximum received signal strength, or the maximum
signal-to-interference-plus-noise ratio (SINR) because the transmit
power amongst HPNs and RRHs are significantly different. Meanwhile,
larger cluster size can provide better SE performances for the
desired UE at the cost of leading to higher fronthual consumption.
Consequently, the RRH/HPN association strategy should optimize the
cluster size for each UE, in which the fronthual overhead and
cooperative gains should be balanced. Further, the association
whether with RRHs or HPN has great impact on SE performances of
H-CRANs. If UEs only associate with RRHs, the proposed H-CRAN is
simplified to the C-RAN. The single nearest and $N$-nearest RRH
association strategies for H-CRANs are tackled
in\textcolor[rgb]{1.00,0.00,0.00}{\cite{peng_RRH}}. As shown in Fig.
\ref{PC:b}, the ergodic capacity performances under different number
of association RRHs with the varying transmit power of RRHs are
compared. The capacity grows monotonically with increasing transmit
power because the interference can be avoided due to the centralized
cloud computing based cooperative processing. The capacity gain of
the 2-nearest RRH association over the single nearest RRH
association is significant. However, the capacity gaps among 4, 8
and infinite RRH association strategies are not large, which
indicates that no more than 4 RRHs are associated for each UE to
balance performance gains and implementation cost.

Besides enhancing SE and EE performances, there is a strong
incentive to improve mobility performances in H-CRANs, e.g., there
are less handover failure, lower Ping-Pong rate and lower drop ratio
for high-mobility UEs than those in C-RAN. Though the handover
between adjacent RRHs is administrated in the same BBU pool, which
means that most handover signallings are avoided, the RRH
re-association and radio resource reconfiguration are still
challenging to the mobility performance. Therefore, in the
high-density C-RANs, high-mobility UEs are likely to be victims who
experience radio link failures (RLFs) before completing the handover
process. Fortunately, in H-CRANs, the high-mobility UEs are served
by HPNs with reliable connections, and the low-mobility UEs are
preferred to access the RRHs. Consequently, by contrast to C-RANs,
there are significant mobility performance gains in H-CRANs.

\section{Promising Key H-CRAN Technologies}

To take full advantages of H-CRANs, the cloud computing based
cooperative processing and networking techniques in PHY, medium
access control (MAC) and network layers should be exploited. The
cloud computing based CoMP (CC-CoMP) technique as the evolution of
the traditional CoMP in 4G systems is utilized to fulfill the
interference cancelations and collaboration amongst RRHs and HPNs.
The large-scale cooperative multiple antenna (LS-CMA) technique with
the large-scale antenna array is adopted to provide additional
diversity and multiplexing gains for HPNs. Through CC-CRRM, the
radio resources amongst RRHs can be shared and virtualized, and the
cross-interference between RRHs and HPNs can be coordinated.
Comparing with HetNets and C-RANs, H-CRAN is more complex and costly
in network planning and maintaining. Therefore, the cloud computing
based self-organizing network (CC-SON) is indispensable to enhance
intelligences and lower human costs.

\subsection{Cloud Computing based Coordinated Multi-Point (CC-CoMP)}

\begin{figure}[!htp]
\centering
\includegraphics[width=6.0in]{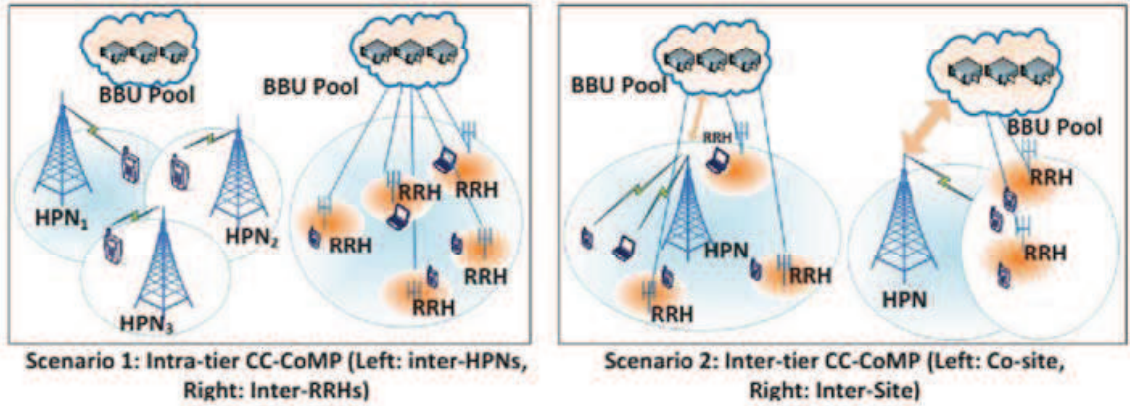}
\caption{\textbf{Typical Scenarios for CC-CoMP in
H-CRANs}}\label{LSCoMP}\vspace*{-1em}
\end{figure}

The potential applications of CC-CoMP in H-CRANs exist homogenous
and heterogeneous scenarios, which are denoted as intra-tier and
inter-tier, respectively. In the intra-tier CC-CoMP shown as
scenario 1 in Fig. \ref{LSCoMP}, nodes in the same tier, i.e.,
amongst HPNs or amongst RRHs, are required to transmit coordinately.
Since all RRHs in H-CRANs are connected with the BBU pool, such
collaboration can be realized through virtual beamforming, where the
beamers can be formed at the BBU pool. Moreover, the inter-tier
CC-CoMP shown as scenario 2 in Fig. \ref{LSCoMP} relies on the
collaborations between RRHs and HPNs, and such cross-tier
collaborations may put a heavy burden of computational complexity on
the BBU pool. Therefore, to ease the burden of BBU pool, distributed
schemes in the centralized H-CRAN structure could be partially
executed instead.

To mitigate intra-tier and inter-tier interference efficiently, the
traditional CoMP in 4G should work in perfect and ideal status to
achieve significant cooperative processing gains, which arouses high
complexities, increased synchronization requirements, complicated
channel estimation efforts, and huge signaling overhead.
Fortunately, as the evolution of CoMP, the CC-CoMP relies on the
large-scale spatial cooperative processing in the centralized BBU
pool, in which most challenges of CoMP are alleviated. The biggest
challenging for CC-CoMP is that the full-scale coordination requires
the processing of very large channel matrix consisting of channel
coefficients from all UEs to all RRHs/HPNs, leading to high
computational complexity and channel estimation overhead.

Accurately, only a small fraction of the entries in the channel
matrix have reasonably gains because any UE is only close to a small
number of neighbor RRHs/HPNs. Consequently, the sparsity or
near-sparsity channel matrices can be utilized during designing the
efficient CC-COMP, where the core problem is to what extent the
channel matrix for CC-CoMP can be compressed without substantially
compromising the system performance. On one hand, the improvement of
SE performances might be very limited when the scale size of CC-CoMP
is small. On other hand, by setting the participants overly, the
exchanging signaling overhead and their channel state information
(CSI) amount are intractable. Further, the accuracy and
instantaneity of CSI decline with the increasing scale size, which
degrades SE performances severely. Therefore, it is critical to
clarify the boundary conditions in which CC-CoMP can achieve
significant gains in H-CRANs. To decrease the implementation
complexity, a joint RRH selection and power minimization beamforming
problem was proposed
in\textcolor[rgb]{1.00,0.00,0.00}{\cite{Letaief}}, where a
bi-section group sparse beamforming (GSBF) algorithm and an
iterative GSBF algorithm are proposed. It is demonstrated that the
proposed bi-section GSBF algorithm is a better option if the number
of RRHs is huge due to its low complexity, while the iterative GSBF
algorithm can be applied to provide better performances in a
medium-size network.

Considering the superior centralized processing capability of H-CRAN
provides great conveniences for making CC-CoMP work efficiently, the
combinations with other advanced techniques, such as interference
alignment, should be researched in the future.

\subsection{Large-Scale Cooperative Multiple Antenna Processing (LS-CMA)}

The LS-CMA technique, also known as massive MIMO, is equipped with
hundreds of low-power antennas at a co-located HPN site, which is
presented to improve capacity, extend coverage, and decrease antenna
deployment complexity. Due to the law of large numbers, the channel
propagation condition can be hardened, which can ensure that the
transmission capacity increases linearly as the number of antennas
increases, and EE performances can be improved as well. Compared
with the traditional single antenna configuration, the LS-CMA for
HPNs can increase the capacity 10 times or more and simultaneously
improve the radiated EE performances on the order of 100 times
in\textcolor[rgb]{1.00,0.00,0.00} {\cite{bib:CC-MA}}, where the
100--element linear array is deployed in HPNs, and the ideal
backhaul is assumed.

\begin{figure}[!htp]
\centering
\includegraphics[width=6.0in]{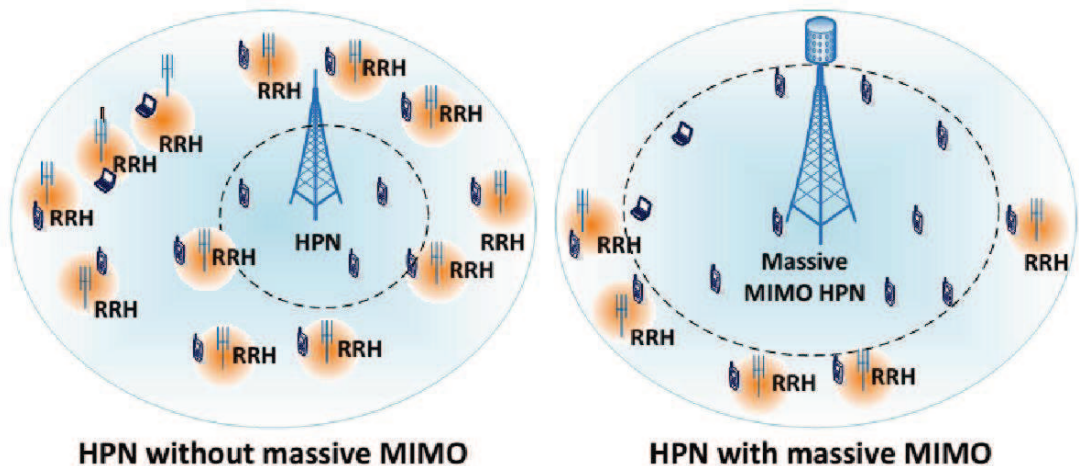}
\caption{\textbf{Typical Scenarios for LS-CMA in H-CRANs}
}\label{MassiveMIMO}\vspace*{-1em}
\end{figure}

Unlike RRHs, HPN needn't to upload all observations to the
centralized BBU pool for the base-band large-scale signal
processing. By using HPNs with LS-CMA, instead of deploying a huge
number of RRHs in some coverage areas, the constrained fronthauls
between RRHs and the BBU pool can be released. Moreover, as shown in
Fig. \ref{MassiveMIMO}, compared to HPNs without LS-CMA, or the
conventional C-RAN scenarios, the HPNs with LS-CMA reduce their
interferences to the adjacent RRHs/HPNs since LS-CMA can serve a
large area, which widens the serving distance and dilutes the
density of active RRHs.

Most of existing works on LS-CMA mainly focus on the PHY layer or
homogeneous scenario. Considering the cloud computing feature in
H-CRANs, the inter-tier CC-CoMP, performing the cooperative
beamforming between HPNs with LS-CMA and RRHs, is an important
approach to mitigate inter-tier interference in H-CRANs, which
results in improving EE and SE performances significantly.
Furthermore, considering that the maximum allowable number of
orthogonal pilot sequences is upper bounded by the duration of the
coherence interval divided by the channel delay-spread, the pilot
contamination as a basic phenomenon is not really specific to
LS-CMA, but its impact on LS-CMA appears to be much more profound
than that on the classical MIMO. In\textcolor[rgb]{1.00,0.00,0.00}
{\cite{bib:CC-MA_pilot}}, the Bayesian channel estimation method
making explicit use of covariance information in the inter-cell
interference scenario with pilot contamination was developed, which
leads to a complete removal of pilot contamination effects in the
case that covariance matrices satisfy a certain non-overlapping
condition on their dominant subspaces.

Too many HPNs with LS-CMA in H-CRANs sacrifices performance gains of
RRHs, while too few HPNs make H-CRANs change into C-RANs. Therefore,
how to achieve the best tradeoff between HPNs and RRHs is critical
to make H-CRANs work efficiently. The optimized densities and
deployment sites of HPNs and RRHs should be researched in the
future.

\subsection{Cloud Computing based Cooperative Radio Resource Management (CC-CRRM)}

To fully unleash the potential advantages of H-CRAN, the intelligent
CC-CRRM is urgent and there are various technical challenges
involved. First, CC-CRRM needs to support real-time and bursty
mobile data traffic, such as mobile gaming, vehicle-to-vehicle
communications and high-definition video streaming applications.
Therefore, CC-CRRM should have the time delay aware ability. Most of
the traditional RRMs are based on heuristics and there is lack of
theoretical understanding on how to design delay-aware CC-CRRM.
Second, the CC-CRRM has to be scalable with respect to H-CRAN size,
while the traditional RRMs are infeasible due to the huge
computational complexity as well as signaling latency / complexity
involved. These challenges become even worst for H-CRANs because
there are more thin RRHs connected to the BBU pool through the
constrained fronthaul. Unlike conventional cross-layer RRM, which is
designed to optimize the resource of a single base station, the
CC-CRRM involves shared radio resources among all RRHs/HPNs, and
thus the scalability in terms of computation and signaling is a key
obstacle\textcolor[rgb]{1.00,0.00,0.00}{\cite{bib:CRRM}}.

\begin{figure}[!htp]
\centering
\includegraphics[width=4.5in]{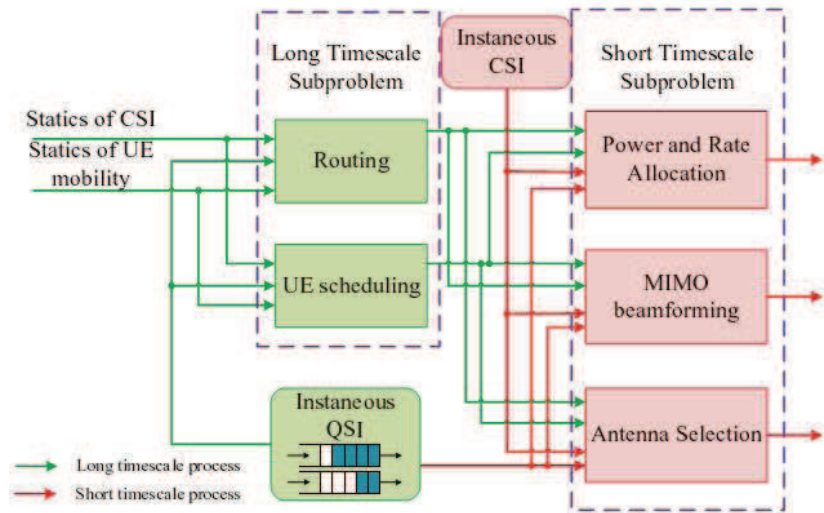}
\caption{\textbf{Decomposition of Mixed Timescale Stochastic
Optimization for CC-CRRM} }\label{CRRM}\vspace*{-1em}
\end{figure}

To tackle the above issues and make H-CRANs practical advancement,
the CC-CRRM should have delay-aware and cross-layer characters.
Accurately, the CC-CRRM for H-CRAN can be regarded as a cloud
computing based stochastic optimization problem, which adapts the
radio resources (such as power, data rate, CC-CoMP/LS-CMA, user
scheduling, and RRH/HPN association) according to the real-time CSI
and queue state information (QSI). In practice, the signallings and
controls are usually enforced at the frame level in the PHY or lower
MAC layers, while they are usually done at longer timescales in the
upper MAC and network layers. Based on the structural property of
the stochastic control problem and the separation of timescales, the
stochastic control problem for H-CRANs can be decomposed into a
number of lower dimension subproblems and further be solved by
stochastic online learning techniques, as depicted in Fig.
\ref{CRRM}.

The delay-aware CC-CRRM for H-CRAN is adaptive to both the global
QSI and CSI, which not only captures the opportunity to transmit
indicated by the QSI, but also captures the urgency of data flows
indicated by the CSI. With the separation of timescales, the CC-CRRM
requires the reduced signaling overhead and computational
complexity. The stochastic online learning instead of heuristic
methods guarantees that the CC-CRRM solution is robust to
uncertainty in CSI estimation, traffic bursty arrival statistics as
well as other key parameters. However, since the delay-aware CC-CRRM
is based on the global QSI and CSI, the underlying
curse-of-dimensionality associated with the system states and
coupled queue dynamics will complicate the derivation of a scalable
CC-CRRM in H-CRANs. Fortunately, the utilization of stochastic
differential equation in Markov decision process (MDP) lights a new
way to facilitate the derivation of low complexity and scalable
policy for the H-CRAN and has attracted further
studies\textcolor[rgb]{1.00,0.00,0.00}{\cite{bib:CRRM2}}.

\subsection{Cloud Computing based Self-Organizing H-CRANs (CC-SON)}

In the network layer of H-CRANs, the self-organizing functionnaires
are critical to guarantee the giant RRHs and HPNs work with high SE
and EE performances. Self-organizing network (SON) technology is
able to minimize human interventions in networking processes, which
was proposed to reduce operational costs for service providers in
LTE cellular systems and
HetNets\textcolor[rgb]{1.00,0.00,0.00}{\cite{bib:SON2}}. Considering
a huge number of parameters should be configured and optimized, the
topology is dynamical due to switching on/off RRHs adaptively, and
the radio resources are shared and cooperatively processed, CC-SON
in H-CRANs is the key to integrate network planning, configuration,
and optimization into a unified automatic process requiring minimal
manual interventions with the centralization of cloud computing.
CC-SON allows operators to streamline their operations, not only
reducing the complexity of managing co-channel interference in
H-CRANs, but also saving operational costs to all RRHs and HPNs.
CC-SON is used to harmonize the whole network management approaches
and improve the overall operational efficiency. On the other hand,
the availability of CC-SON solutions leads to identify powerful
optimization strategies, mitigate co-channel interferences and
improve EE performances.

Different from the hybrid SON architecture in HetNets, due to the
existence of centralized BBU pool, the self-configuration,
self-optimization, and self-healing functionalities in CC-SON are
implemented in the hierarchical SON architecture, in which the
centralized structure is utilized in the BBU pool for all RRHs, and
the distributed structure is adopted between the BBU pool and HPNs.
For the self-configuration, since RRHs are utilized to provide a
high capacity transparently, the physical cell identifier (PCI)
assignment is not necessary for RRHs, and only the radio resource
should be self-assigned to each RRHs intelligently. However, both
self-configurations of PCI and radio resource should be fulfilled in
HPNs. These self-configuration cases for HPNs should be handled with
the help of the centralized BBU pool. For the self-optimization of
H-CRANs, energy saving and mobility robustness optimization are two
key cases, in which RRHs should be turned on/off automatically, and
the overlap coverage of RRHs and HPNs should be intelligently
compensated with each other. For the self-healing, the outage
detection and performance compensation are executed mainly in the
BBU pool, and the shared radio resources among RRHs can be
adaptively re-configured to compensate the outage coverage. Note
that these aforementioned SON cases are triggered by the
minimization of drive test mechanisms. The specified CC-SON related
algorithms should be researched further for H-CRANs in the future.

\section{Challenges and Open Issues in H-CRANs}

Although there have been some progresses and initial achievements in
the aforementioned potential system architecture, performance
analysis, key techniques for H-CRANs, there are still many
challenges ahead, such as the theoretical performance analysis with
stochastic geometry, optimal resource allocation with the
constrained fronthaul, and standard development. Some classical
challenges as samples are discussed in this section.

\subsection{Performance Analysis with Stochastic Geometry}

With the cellular network evolving from conventional grid cellular
network to HetNets, it is highly desirable to study the capacity of
HetNets, where locations of LPNs and UEs follow with the stationary
Poisson point process (PPP). Owing to its tractability, PPP model
leads to simple closed-form expressions for key metrics such as
coverage probability and average rate over the entire
network\textcolor[rgb]{1.00,0.00,0.00}{\cite{bib:Andrews}}.
Considering the capacity scales with the limited fronthaul, the
optimal rate allocation that maximizes overall sum rates in C-RAN
was analyzed
in\textcolor[rgb]{1.00,0.00,0.00}{\cite{bib:CRANcapacity}}, in which
the fronthaul link rates are demonstrated to scale logarithmically
with the SINR at each node.

As a tractable tool of modeling interference, stochastic geometry
can be utilized to capture the characteristic of the locations of
RRHs/HPNs and hence develop accurate performance results, e.g.
coverage probability, average capacity and area spectral efficiency.
However, the differences between the H-CRANs and the conventional
HetNets and C-RANs in utilizing the stochastic hemometry are
twofold. First, interference coordination is much more feasible due
to the centralized computation in the BBU pool for H-CRAN than that
for HetNet. Interference cancelation and collaboration in the
spatial domain is confined to HetNets due to the constrained
cross-tier backhaul, whereas it is possible to perform more
efficient large-scale cooperative signal processing algorithms in
H-CRANs, e.g. CC-CoMP and LS-CMA, to achieve higher performance
gains. Based on the stochastic geometry, it is significant to
propose a tractable yet practical strategy to derive the analytical
closed-form expressions for SE and EE performances. Based on these
expressions, the impact of imperfect fronthaul link can be evaluated
and near sparse beamforming schemes can be designed jointly with the
RRH selection and power allocation. Second, H-CRANs can obtain
additional performance gains over C-RAN from the mobility
enhancement. A tractable mobility model should be constructed to
analytically evaluate performance gains in terms of handover success
ratio and sojourn time.

A general framework of performance metrics with stochastic geometry
should be developed in the future to derive theoretical SE, EE and
mobility performances of H-CRANs. Particularly, the closed-form SE
expression with stochastic geometry for C-RANs is still not derived,
which is the baseline for investigating the theoretical performance
of H-CRANs. The EE and mobility theoretical performances with
stochastic geometry of H-CRANs are still open issues and more
challenging than the SE performance analysis because they make the
non-convex optimization problem more complex.

\subsection{Performance Optimization of Constrained Fronthaul}

The non-ideal fronthaul links between RRHs and the BBU pool
deteriorate overall SE and EE performances of H-CRANs, where both
the limited capacity and time latency constraints are inevitable in
practice. Besides, when the network scale grows larger, it's
impractical to assume that all ideal CSIs of the entire network are
tractable.

The constraints of the fronthaul affect SE performances in several
ways. The total transmission bit rate per RRH should not be larger
than the capacity of its corresponding fronthaul link. The
insufficient fronthaul capacity prevents RRH to make full use of
available radio radio resources. Such problem becomes even more
severe when CC-CoMP, LS-CMA and CC-CRRM are utilized. Besides, the
time latency of the fronthaul also plays a key handicap to optimize
both SE and EE performances. Most obviously, the time latency make
CSI outdated, which results in deteriorating the accuracy of sparse
beamforming.

To overcome these constrained fronthaul related problems, advanced
cooperative processing and resource allocation strategies need to be
researched. The optimal resource allocation solutions to maximize SE
or EE under the constrained fronthaul should be proposed, which is a
non-deterministic polynomial-time (NP)-complete problem in general.

\subsection{H-CRAN Standardizations}

The standardizations of H-CRANs should be strictly backward
compatible with both C-RANs and HetNets, which have been widely
discussed in 3GPP\textcolor[rgb]{1.00,0.00,0.00}{\cite{bib:3GPP}}.
The standards of small cell enhancement in 3GPP Release 12 include
higher order modulation, almost blank subframe, small cell on/off,
SON, energy saving, and CoMP operation in HetNets with non-ideal
backhaul. As the extension and evolution of HetNets, the RRH on/off,
CC-CoMP, LS-CMA, CC-CRRM, and CC-SON could be standardized for
H-CRANs, which can be considered as potential issues in 3GPP Release
13 and beyond. The functionalities and interfaces of backhaul links
have been standardized for HetNets in 3GPP to achieve the inter-cell
and inter-tier CoMP gains. These standard results can be extend
directly to standardizations of the backhaul link between the BBU
pool and HPNs in H-CRANs. Unfortunately, standardizations of
fronthual links between RRHs and the BBU pool are still not
straightforward, which should be emphasized in the future standard
works.

To make HPNs coordinate with the BBU pool more timely, flexibly and
integrally, the functionalities of HPNs, RRHs, and the BBU pool, as
well as the air interfaces of backhaul and fronthaul links for
H-CRANs should be highlighted in the future 3GPP standards. First,
the X2/S1 interfaces for the backhaul link between the BBU pool and
HPNs should be enhanced from the existing X2/S1 interfaces for
HetNets, which are mainly designed for CoMP to support fast adaptive
interference coordinations in time/frequency domain and joint
processing in spatial domain. Second, radio resource management,
such as resource allocation, RRH/HPN association and power control,
should be partially performed by HPNs instead of totally executed at
the BBU pool. Meanwhile, the CC-SON protocol should be enhanced from
current SON standards for HetNets.

\section{Conclusion}

In this article, we have provided a summary of recent advancements
in the computing convergence of heterogeneous wireless networks. The
heterogeneous cloud radio access network (H-CRAN) is proposed as a
promising new paradigm to achieve high SE and EE performances
through the combination of cloud computing and HetNets. The system
architecture, performance analysis, cloud computing based
cooperative processing and networking techniques for the proposed
H-CRANs have been surveyed. In particular, the key large-scale
cooperative processing and networking techniques including cloud
computing based CoMP, CRRM, SON have been briefly summarized.
Further, potential challenges and open issues in H-CRANs have been
discussed as well. The presented key techniques and potential
solutions for H-CRANs provide breakthroughs of theories and
technologies for the advanced next generation wireless communication
systems.

\begin{IEEEbiography}[{\includegraphics[width=1in,height=1.25in]{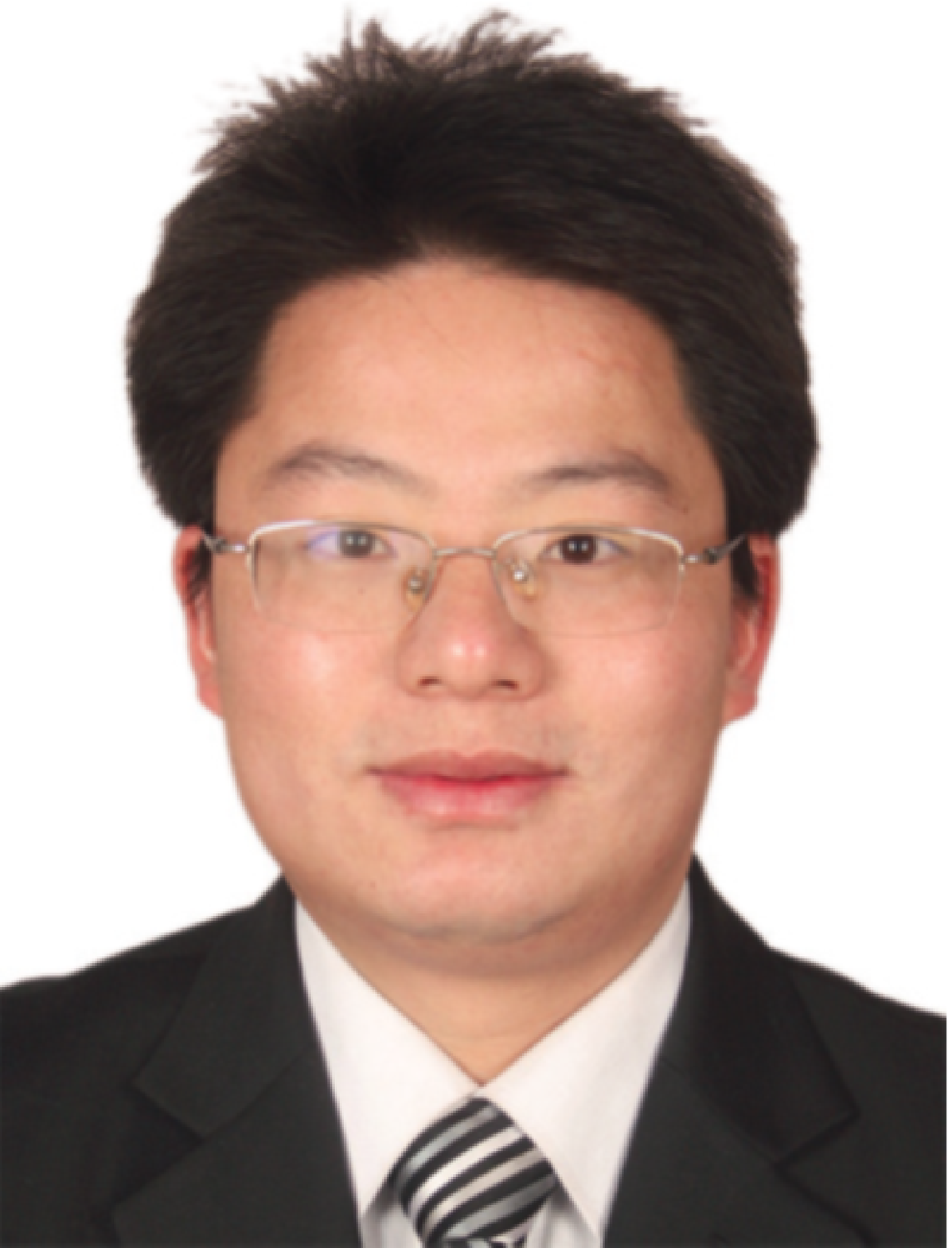}}]{Mugen Peng}
(M'05--SM'11) received the B.E. degree in Electronics Engineering
from Nanjing University of Posts \& Telecommunications, China in
2000 and a PhD degree in Communication and Information System from
the Beijing University of Posts \& Telecommunications (BUPT), China
in 2005. After the PhD graduation, he joined in BUPT, and has become
a full professor with the school of information and communication
engineering in BUPT since Oct. 2012. During 2014, he is also an
academic visiting fellow in Princeton University, USA. He is leading
a research group focusing on wireless transmission and networking
technologies in the Key Laboratory of Universal Wireless
Communications (Ministry of Education) at BUPT, China. His main
research areas include wireless communication theory, radio signal
processing and convex optimizations, with particular interests in
cooperative communication, radio network coding, self-organization
networking, heterogeneous networking, and cloud communication. He
has authored/coauthored over 40 refereed IEEE journal papers and
over 200 conference proceeding papers.

Dr. Peng is currently on the Editorial/Associate Editorial Board of
IEEE Access, International Journal of Antennas and Propagation
(IJAP), China Communication, and International Journal of
Communications System (IJCS). He has been the guest leading editor
for the special issues in IEEE Wireless Communications, IJAP and the
International Journal of Distributed Sensor Net- works (IJDSN). He
was the guest editor of IET Communications. He is serving as the
track chair for GameNets 2014, WCSP 2013 and ICCT 2011, the workshop
co-chair of ChinaCom 2012, and the leading co-chair for So-HetNets
in IEEE WCNC 2014, SON-HetNet 2013 in IEEE PIMRC 2013, and SON 2013
in ChinaCom 2012.

Dr. Peng was honored with the Best Paper Award in ICCTA 2011,
IC-BNMT 2010, and IET CCWMC 2009. He was awarded the first Grade
Award of Technological Invention Award in Ministry of Education of
China for his excellent research work on the hierarchical
cooperative communication theory and technologies, and the Second
Grade Award of Scientific \& Technical Progress from China Institute
of Communications for his excellent research work on the
co-existence of multi-radio access networks and the 3G spectrum
management in China.
\end{IEEEbiography}

\begin{IEEEbiography}[{\includegraphics[width=1in,height=1.25in]{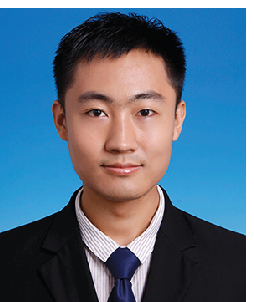}}]{Yuan Li}
received the B.E. degree in communication engineering from Shandong
University, Jinan, China, in 2009, and Ph.D. degree in Communication
and Information System from Beijing University of Posts and
Telecommunications in 2014. He is currently working at Huawei
Technologies Co. Ltd. His current research interests include
interference coordination, multiple antenna technology, small cell
and heterogeneous networks.
\end{IEEEbiography}

\begin{IEEEbiography}[{\includegraphics[width=1in,height=1.25in]{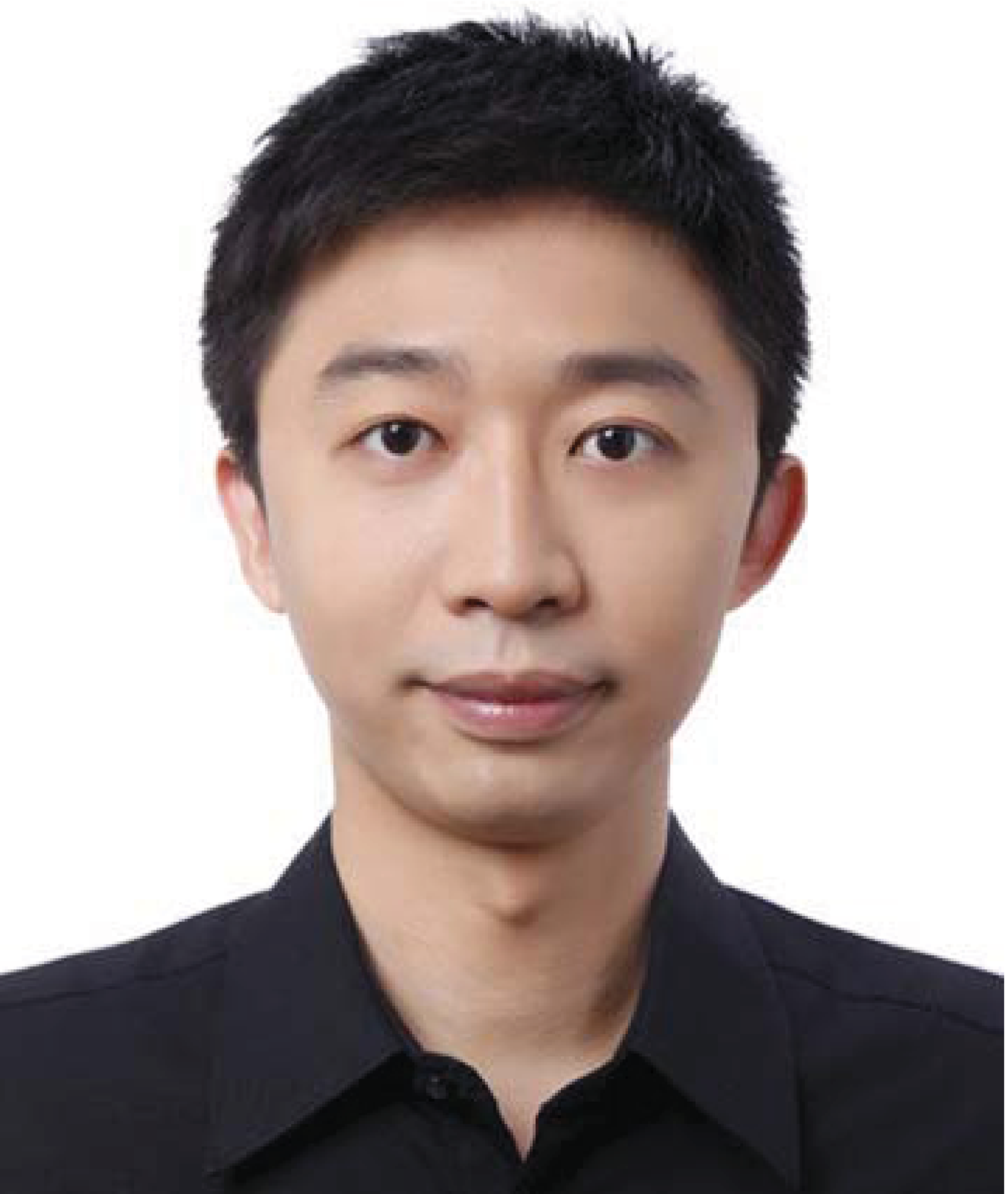}}]{Jiamo Jiang}
received the Ph.D. degree in Communication and Information System
from the Beijing University of Posts and Telecommunications,
Beijing, China, in 2014. He is currently with Institute of
Communication Standards Research, China Academy of Telecommunication
Research of MIIT, Beijing, China. His research focuses on
self-organizing networks and energy-efficient resource allocation in
heterogeneous networks.
\end{IEEEbiography}

\begin{IEEEbiography}[{\includegraphics[width=1in,height=1.25in]{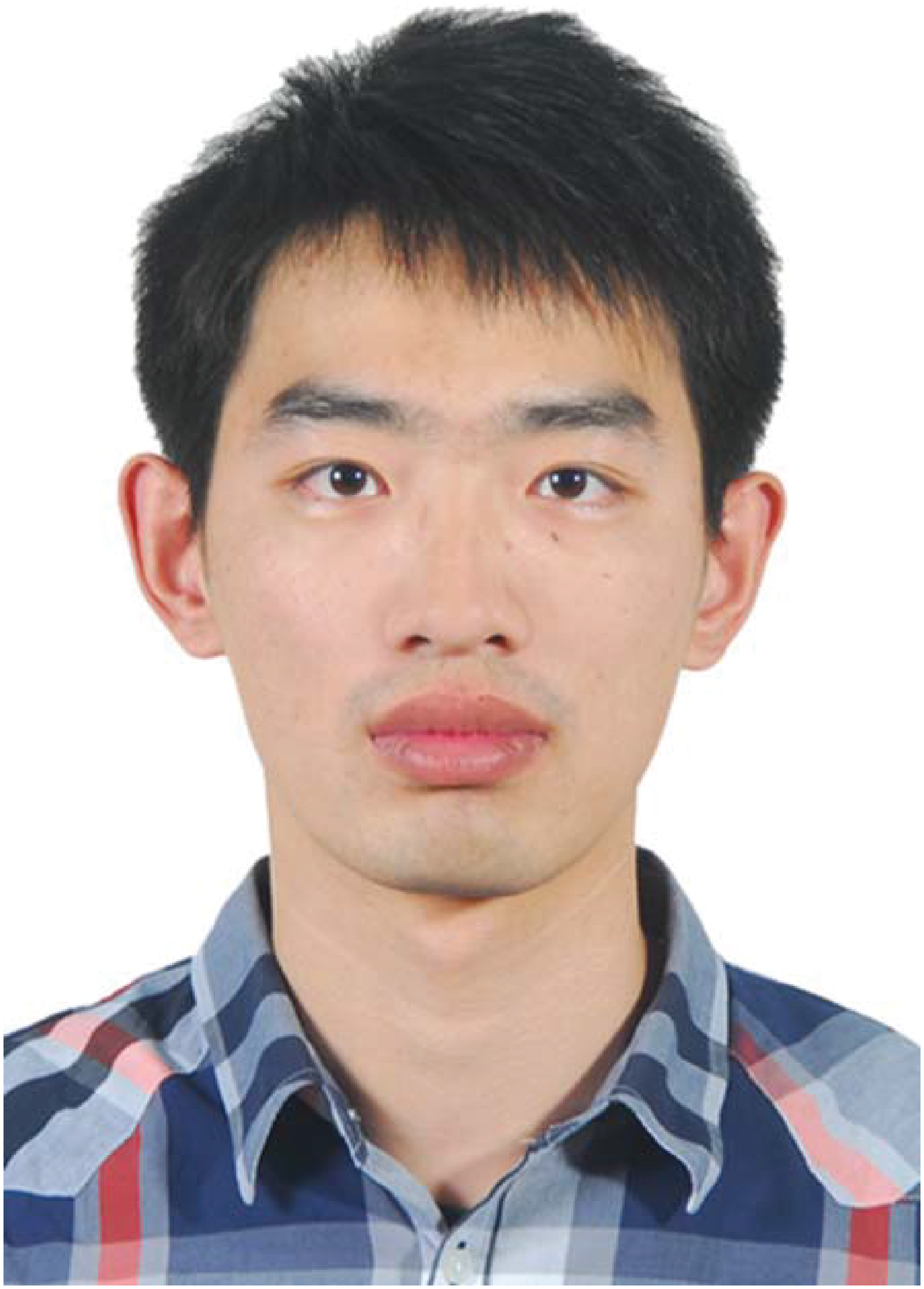}}]{Jian Li}
received his B.E. degree in Measuring and Control Technology and
Instrumentations from Nanjing University of Posts and
Telecommunications, Nanjing, China, in 2010. He is currently
pursuing his Ph.D. degree at the key laboratory of universal
wireless communication (Ministry of Education) in Beijing University
of Posts and Telecommunications (BUPT), Beijing, China. His current
research interests include delay-aware cross-layer radio resource
optimization for heterogeneous networks (HetNets) and heterogeneous
cloud radio access networks (H-CRANs).
\end{IEEEbiography}

\begin{IEEEbiography}[{\includegraphics[width=1in,height=1.25in]{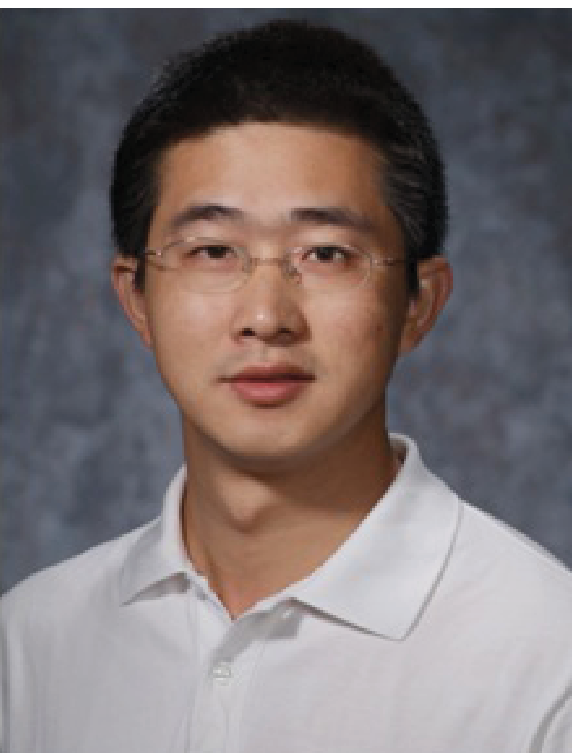}}]{Chonggang Wang}
(SM'09) received the Ph.D. degree from Beijing University of Posts
and Telecommunications (BUPT) in 2002. He is currently a Member of
Technical Staff with InterDigital Communications. His R\&D focuses
on: Internet of Things (IoT), Machine-to-Machine (M2M)
communications, Heterogeneous Networks, and Future Internet,
including technology development and standardization. He
(co-)authored more than 100 journal/conference articles and book
chapters. He is on the editorial board for several journals
including IEEE Communications Magazine, IEEE Wireless Communications
Magazine and IEEE Transactions on Network and Service Management. He
is the founding Editor-in-Chief of IEEE Internet of Things Journal.
He is serving and served in the organization committee for
conferences/workshops including IEEE WCNC 2013, IEEE INFOCOM 2012,
IEEE Globecom 2010-2012, IEEE CCNC 2012, and IEEE SmartGridComm
2012. He has also served as a TPC member for numerous conferences
such as IEEE ICNP (2010-2011), IEEE INFOCOM (2008-2014), IEEE
GLOBECOM (2006-2014), IEEE ICC (2007-2013), IEEE WCNC (2008-2012)
and IEEE PIMRC (2012-2013). He is a co-recipient of National Award
for Science and Technology Achievement in Telecommunications in 2004
on IP QoS from China Institute of Communications. He received
Outstanding Leadership Award from IEEE GLOBECOM 2010 and
InterDigital's 2012 and 2013 Innovation Award. He served as an NSF
panelist in wireless networks in 2012. He is a senior member of the
IEEE and the vice-chair of IEEE ComSoc Multimedia Technical
Committee (MMTC) (2012-2014).
\end{IEEEbiography}

\end{document}